\documentclass[12pt,a4paper]{article}

\usepackage{epsfig}
\usepackage{amsmath,amsfonts,amssymb}
\usepackage{t1enc}
\usepackage{cite}

\providecommand{\openone}{\leavevmode\hbox{\small1\kern-3.8pt\normalsize1}}

\begin{document}

\vspace*{-2cm}
\begin{flushright}
UG--FT--168/04 \\
CAFPE--38/04 \\
ROMA-1401/05 \\
hep-ph/0502189
\end{flushright}

\begin{center}
\begin{Large}
{\bf Flavour and polarisation in heavy neutrino \\[0.2cm]
production  at $\boldsymbol{e^+e^-}$ colliders}
\end{Large}

\vspace{0.5cm}
F. del Aguila $^a$, J. A. Aguilar-Saavedra $^b$, A. Mart\'{\i}nez de la Ossa
$^a$ and D. Meloni $^c$ \\[0.2cm]
{\it $^a$ Departamento de Física Te\'orica y del Cosmos and Centro Andaluz de
Física de Partículas elementales (CAFPE), \\
Universidad de Granada, E-18071 Granada, Spain} \\[0.1cm]
{\it $^b$ Departamento de Física and CFTP, \\
  Instituto Superior Técnico, P-1049-001 Lisboa, Portugal} \\[0.1cm]
{\it $^c$ INFN Sezione di Roma I and Dip. di Fisica, \\
Univ. di Roma ``La Sapienza'' P. le A. Moro 2, I-00185 Rome, Italy} \\
\end{center}

\begin{abstract}
We analyse $\ell W \nu$ production at ILC, paying special attention to the role
of the final lepton flavour and beam polarisation in the search for a new heavy
neutrino $N$.  We show that a sizeable coupling to the electron $V_{eN} \sim
10^{-2}$ is necessary  to have an observable signal in any of the channels,
despite the fact that the signal may be more visible in muon or tau final
states. The non-observation of a heavy neutrino at ILC will improve the present
upper bound on its mixing with the electron by more than one order 
of magnitude, $V_{eN} \leq 0.007$ for $m_N$  between 200 and 400 GeV.
\end{abstract}

\section{Introduction}

A 500 GeV $e^+ e^-$ International Linear Collider (ILC) offers a clean
environment for the study of physics beyond the Standard Model (SM) at a scale
of few hundreds of GeV.
Its potential is not limited to the study of low energy supersymmetry and
precision top quark physics \cite{tdr}. On the contrary, such a machine is a
helpful tool
for the investigation of less conventional models, and it might even reveal
unexpected new physics.

In this last category might be classified the possible existence of heavy
neutrinos with masses of few hundreds of GeV. They are absent in the simplest
SM extensions, as long as they do not provide a light neutrino mass generation
mechanism,\footnote{In principle they can give see-saw type contributions
to light neutrino masses \cite{seesaw}, but these contributions of the order
$V_{eN}^2 m_N \sim 10^7\ \mathrm{eV}$ would be too large compared to the
typical neutrino mass size $m_\nu \sim 1$ eV. This means that some symmetry or
accidental cancellation is required to reproduce the observed  neutrino masses
\cite{G}.} nor an explanation of the observed baryon to photon ratio of the
universe,\footnote{Neutrino singlets with large Majorana masses can produce an
excess of lepton number $L$ which can be converted into the observed
baryon asymmetry through $B + L$ violating sphaleron interactions \cite{FY}.
Nevertheless, the heavy neutrinos which may generate a lepton asymmetry
large enough  have a small mixing with the light fermions, typically of order
$\sqrt{\tilde m_\nu / m_N} \lesssim 10^{-8}$, with $\tilde m_\nu$ the effective
light neutrino mass relevant to the process \cite{BDBP}, being then their
production rates negligible.}
$\eta_B = 6.5 \times 10^{-10}$ \cite{bau}.
However, they appear in Grand Unified Theories, in particular those based on
$\mathrm{SO}(10)$ and larger groups like $\mathrm{E}_6$ \cite{GUT}, and they may
acquire masses much smaller  than the unification mass scale \cite{BR}.
Kaluza-Klein towers of neutrinos  are also predicted by models with large extra
dimensions and bulk fermions \cite{EDBF}, being possible to have the lightest
heavy modes near the electroweak scale \cite{AS}. Their presence is not
experimentally excluded and, if they exist and their mixing with the electron is
$O(10^{-2})$, they will be produced at ILC. Conversely, if they are not
observed, present bounds on their mixing with the electron will be improved
by one order of magnitude.

In $e^+ e^-$ annihilation the most favourable process for the observation of
heavy neutrino singlets $N$ is $e^+ e^- \to N \nu \to \ell W \nu$. We will
refer to it
as $\ell W \nu$ production from now on. This process has been studied by
different authors \cite{ALZ,D,GZ3,ACMV}. Here we review previous work including
the SM background as well as the effects of initial state radiation (ISR) and
beamstrahlung, which have a great impact in the observability of the heavy
neutrino, {\em e.g.} for $m_N = 300$ GeV they reduce the signal to background
ratio by more than a factor of two. In addition, we examine the role of flavour
and beam polarisation in the search of heavy neutrinos. As we will show in this
Letter, an $eNW$ coupling $O(10^{-2})$ is necessary to produce a heavy neutrino
at a detectable level in any of the channels. In this
situation, the use of beam polarisations $P_{e^-} = -0.8$, $P_{e^+} = 0.6$
improves the statistical significance of the signal, which may be more visible
in final states with $\ell = \mu,\tau$. In case that $N$ does not couple to the
electron, the opposite polarisations $P_{e^-} = 0.8$, $P_{e^+} = -0.6$ can be
used to enhance the signal and reduce the background. However, they do not
suffice to make the signal observable unless very large integrated luminosities
are collected. The Dirac or Majorana character of $N$ has a negligible effect on
its production cross section and hence does not influence the ILC discovery
potential for a non-decoupled heavy neutrino. Thus, we restrict ourselves to the
case of a Majorana neutrino.

In the following we fix the notation and review present limits on heavy
neutrino masses and mixings, specifying also the SM extensions we will
consider. Then we discuss in turn the main contributions to the signal,
emphasising the phenomenological implications of the final lepton flavour and
beam polarisation. After describing the event generation, we obtain the heavy
neutrino discovery limits at ILC. In the conclusions we summarise our results
and comment on the expected reach of other future experiments.

\section{Heavy neutrino mixing with the light leptons}
\label{sec:2}

Let us first review some well-known results \cite{MBLS} to make explicit our
hipotheses and notation. We assume that besides the three weak isospin
$T_3 = 1/2$ fields $\nu'_{iL}$ there are three neutrino singlets $N'_{Ri}$,
$i=1,2,3$. The neutrino mass term is
\begin{equation}
\mathcal{L}_M = - \frac{1}{2} \,
\left(\bar \nu'_L \; \bar N'_L \right)
\left( \! \begin{array}{cc}
M_L &   \frac{v}{\sqrt 2} Y \\ \frac{v}{\sqrt 2} Y^T & M_R
\end{array} \! \right) \,
\left( \!\! \begin{array}{c} \nu'_R \\ N'_R \end{array} \!\! \right) \,,
\end{equation}
where $\nu'_{iR} \equiv (\nu'_{iL})^c$, $N'_{iL} \equiv (N'_{iR})^c$ and
$Y$, $M_L$, $M_R$ are $3 \times 3$ matrices. The $6 \times 6$ mass
matrix $\mathcal{M}$ can be diagonalised by a unitary matrix,
$\mathcal{U}^\dagger \mathcal{M} \, \mathcal{U}^* = \mathcal{M}_\mathrm{diag}$,
with the mass eigenstates $\nu,N$ related to the weak interaction eigenstates
$\nu',N'$ by
\begin{equation}
\left( \!\! \begin{array}{c} \nu'_L \\ N'_L \end{array} \!\! \right) =
\mathcal{U}
\left( \!\! \begin{array}{c} \nu_L \\ N_L \end{array} \!\! \right) \,,
\quad \quad 
\left( \!\! \begin{array}{c} \nu'_R \\ N'_R \end{array} \!\! \right) =
\mathcal{U}^*
\left( \!\! \begin{array}{c} \nu_R \\ N_R \end{array} \!\! \right) \,.
\end{equation}
The charged lepton mass matrix can be assumed diagonal without loss of
generality. The $6 \times 6$ matrix $\mathcal{U}$ can be written as
\begin{equation}
\mathcal{U} = \left( \! \begin{array}{cc}
V^{(\nu)} & V^{(N)} \\ V^{'(\nu)} & V^{'(N)}
\end{array} \! \right) \,,
\end{equation}
with $V^{(\nu)}$ ($V^{'(N)}$) describing the mixing between the light (heavy)
neutrinos and $V^{(N)}$, $V^{'(\nu)}$ parameterising the light-heavy neutrino
mixing. With this ordering
the extended Maki-Nakagawa-Sakata 
(MNS) matrix \cite{PMNS} 
$V = (V^{(\nu)} ~ V^{(N)} )$ parameterises the charged
and neutral current gauge interactions,
\begin{eqnarray}
\label{ec:lagrW}
\mathcal{L}^W_{CC} & = & - \frac{g}{\sqrt{2}}
\; \bar l_L \gamma^\mu \; V 
\left( \!\! \begin{array}{c} \nu_L \\ N_L \end{array} \!\! \right)
\; W_\mu^- + \mathrm{H.c.} \,, \\
\label{ec:lagrZ}
\mathcal{L}^Z_{NC}\ & = & -
\frac {g}{2 \cos \theta_W} 
\left(\bar \nu_L \; \bar N_L \right) \gamma^\mu  \; X 
\left( \!\! \begin{array}{c} \nu_L \\ N_L \end{array} \!\! \right)
\;  Z_\mu  \,, 
\end{eqnarray}
where $X = V^\dagger V$. For $V^{(N)} = 0$ the matrix $V^{(\nu)}$ is the 
usual $3 \times 3$ unitary MNS matrix.

The most stringent constraints on neutrino mixing result from tree-level
contributions to processes involving neutrinos as external states like 
$\pi \to \ell \bar \nu$ and $Z \to \nu \bar \nu$, and from new one-loop
contributions to processes with only external charged leptons like $\mu \to e
\gamma$ and $Z \to \ell \bar \ell '$ 
\cite{LL,pil1,NRT,pil2,kagan,bernabeu,illana}.
These processes constrain the quantities
\begin{equation}
\Omega_{\ell \ell'} \equiv \delta_{\ell \ell'} - \sum_{i=1}^3 V_{\ell \nu_i}
V_{\ell' \nu_i}^* = \sum_{i=1}^3 V_{\ell N_i} V_{\ell' N_i}^* \,,
\end{equation}
because in the former case we must sum over the external light neutrinos (which
are not distinguished) and in the latter the sum is over loop contributions. 
The first type of processes in particular tests universality. A global fit to
experimental data gives \cite{kagan}
\begin{equation}
\Omega_{ee} \leq 0.0054 \,, \quad \Omega_{\mu \mu} \leq 0.0096 \,, \quad
\Omega_{\tau \tau} \leq 0.016
\label{eps1}
\end{equation}
with a 90\% confidence level (CL). These limits do not depend on the heavy
neutrino masses and are model-independent to a large extent. They imply that
heavy neutrino mixing with the known charged leptons is very small,
$\sum_i |V_{\ell N_i}|^2 \leq 0.0054$, 0.0096, 0.016 for $\ell = e,\mu,\tau$,
respectively. The bound on $\Omega_{ee}$ also guarantees
that  neutrinoless double beta decay is within experimental limits for the
range of heavy neutrino masses we are interested in (larger than 100 GeV) 
\cite{M}.

The second type of processes, involving flavour-changing
neutral currents (FCNC), get new contributions only at the one loop level 
when the SM is extended only  with neutrino singlets, as in our case. These
contributions, and hence the bounds, depend on the heavy neutrino masses. 
In the limit $m_{N_i} \gg M_W$, they imply \cite{bernabeu}
\begin{equation}
|\Omega_{e \mu}| \leq 0.0001 \,, \quad |\Omega_{e \tau}| \leq 0.01 \,, \quad
|\Omega_{\mu \tau}| \leq 0.01 \,.
\label{eps2}
\end{equation}
Except in the case of the first two families, for which experimental constraints
on lepton flavour violation are rather stringent, these limits are of a similar
size as for the diagonal elements. An important difference, however, is that
(partial) cancellations may operate among heavy neutrino contributions. There
can be cancellations with other new physics contributions as well. In this work
we are interested in determining the ILC discovery potential and the limits on
neutrino masses and mixings which could be eventually established. Then,
we must allow for the
largest possible neutrino mixing and FCNC, although they require cancellations
or fine-tuning. Let us examine in more detail the first bound, which is obtained
from present limits on the $e \mu \gamma$ and $e \mu Z$ vertices. The dominant
terms involving heavy neutrinos are proportional to \cite{bernabeu}
\begin{align} 
& \sum_{i=1}^3 V_{e N_i} V_{\mu N_i}^* \phi(m_{N_i}^2/M_W^2) \,,  \nonumber \\
& \sum_{i,j=1}^3 V_{e N_i} X_{N_i N_j} V_{\mu N_j}^* m_{N_i} m_{N_j} 
f( m_{N_i},m_{N_j} ) \,,
\label{ec:loop}
\end{align}
respectively, where
\begin{eqnarray}
\phi (x) & = & \frac{x \, (1-6x+3x^2+2x^3-6x^2\log x)}{2 \, (1-x)^4} \,, \\
f(x,y) & = & \frac{x \, y \, \log x^2/y^2}{x^2 - y^2} \,.
\end{eqnarray}
Then, in principle it is possible to find non-vanishing values of the mixing
angles $V_{e N_i}$, $V_{\mu N_i}$ so that the two sums are cancelled. We can
distinguish two cases. In the flavour conserving case each heavy neutrino only
mixes with one family and both terms are zero because
$V_{e N_i} V_{\mu N_i}^* = 0$ for any $i$ and $X_{N_i N_j}$ is proportional 
to $\delta _{ij}$. In this case $V_{\ell N_\ell}$ can saturate Eq.~(\ref{eps1}).
If there is flavour violation and the lightest heavy neutrino mixes with the
electron and other charged lepton, a mixing large enough to have an observable
signal at ILC requires some fine-tuning to cancel the two terms. 
For instance, if we assume as in Section \ref{sec:5} that 
$m_{N_1} = 300$ GeV and $V_{e N_1} = 0.052$, $V_{\mu N_1} = 0.069$,
$V_{\tau N_1} = 0.126$, we can saturate Eq.~(\ref{eps1}) and make both sums
in Eqs.~(\ref{ec:loop}) negligible taking $V_{e N_2} = -0.052$,
$V_{e N_3} = 0.004$, $V_{\mu N_2} = 0.062$, $V_{\mu N_3} = 0.031$,  
$V_{\tau N_2} = V_{\tau N_3} = 0$, for $m_{N_2} = 500$ GeV, $m_{N_3} = 6$ TeV.

Since in general the mixing between charged leptons and heavy neutrinos 
is very small, 
the matrix $V^{(\nu)}$ (which corresponds to the first three columns
of $V$) is approximately unitary, up to order $V_{\ell N_i}^2$. Moreover, for
the process under consideration the light neutrino masses can be neglected.
Therefore, we can assume in the following 
$V^{(\nu)} \simeq \openone_{3 \times 3}$ in 
Eq.~(\ref{ec:lagrW}), and $X_{\nu_\ell \nu_{\ell'}} \simeq \delta_{\ell \ell'}$,
$X_{\nu_\ell N_i} \simeq V_{\ell N_i}$ in Eq.~(\ref{ec:lagrZ}).

For the calculation of the $\ell W \nu$ cross section including heavy neutrino
contributions the total width $\Gamma_N$ is needed. The partial widths for
$N$ weak decays are
\begin{eqnarray}
\Gamma(N \to W^+ \ell^-) & = &  \Gamma(N \to W^- \ell^+) \nonumber \\
& = & \frac{g^2}{64 \pi} |V_{\ell N}|^2
\frac{m_N^3}{M_W^2} \left( 1- \frac{M_W^2}{m_N^2} \right) 
\left( 1 + \frac{M_W^2}{m_N^2} - 2 \frac{M_W^4}{m_N^4} \right) \,, \nonumber
\\[0.1cm]
\Gamma(N \to Z \nu_\ell) & = &  \frac{g^2}{64 \pi \cos ^2\theta_W} 
|V_{\ell N}|^2
\frac{m_N^3}{M_Z^2} \left( 1- \frac{M_Z^2}{m_N^2} \right) 
\left( 1 + \frac{M_Z^2}{m_N^2} - 2 \frac{M_Z^4}{m_N^4} \right) \,.
\label{ec:widths}
\end{eqnarray}
We ignore the decays $N \to H \nu_\ell$, which may take place if $m_H < m_N$ 
\cite{GZ3,pil3}.
Including these decays in $\Gamma_N$ slightly decreases the $W \ell$ branching
ratios and hence the final signal cross sections, which must be multiplied by
a factor ranging from $3/4$ (for $m_H \ll m_N$) to 1 (for $m_H \geq m_N$).
Independently of the values of the couplings, the branching ratio for charged
current decays is $2/3$ ($1/2$ if we include scalar decays and $m_H \ll m_N$).

\section{General characteristics of the signal}
\label{sec:3}

\begin{figure}[htb]
\begin{center}
\epsfig{file=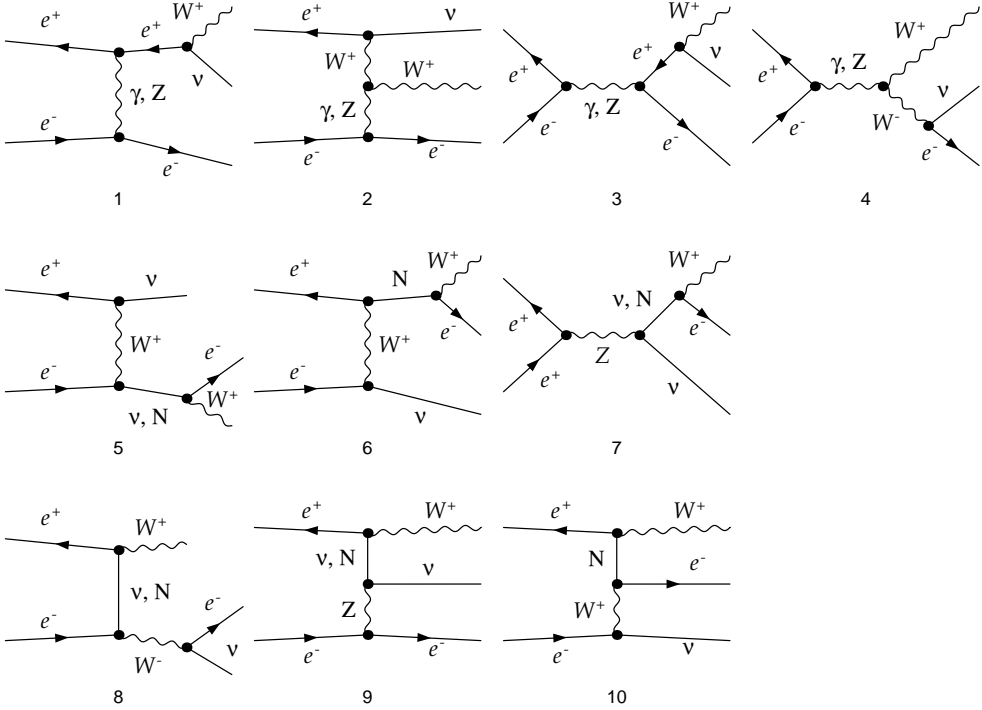,width=13cm,clip=}
\end{center}
\caption{Diagrams contributing to $e^+e^- \to e^- W^+ \nu$. For $\ell =
\mu,\tau$ only diagrams 3-8,10 contribute.}
\label{diag}
\end{figure}

The discovery of a new heavy neutrino in $e^+ e^- \to \ell W \nu$ requires its
observation as a peak in the invariant $\ell W$ mass distribution, otherwise
the irreducible SM background is overwhelming. This requires to reconstruct the
$W$, what justifies to consider $\ell W \nu$ production (instead of general
four fermion production), with $W$ decaying hadronically.
In the evaluation of $e^+ e^- \to \ell^- W^+ \nu$, with $W^+ \to q \bar q'$, 
we will only consider the contributions from the diagrams in Fig.~\ref{diag}, 
neglecting diagrams with four fermions $e^- q \bar q' \nu$ in the final state 
but with $q \bar q'$ not resulting from a $W$ decay. At any rate, we have
checked  that the corresponding contributions are negligible in the phase space
region  of interest.

Let us discuss the contributions and sizes of the different diagrams for  
$e^+ e^- \to \ell^- W^+ \nu$ in Fig. \ref{diag} and what  we can learn from
this type of processes at ILC. 
The first four diagrams are SM contributions. Diagrams 
5, $7-9$ are present within the SM, mediated by a light neutrino, but
they can also involve a heavy one. Diagrams 6 and 10 are exclusive to Majorana
neutrino exchange. The SM contribution has a substantial part from resonant $W^+
W^-$ production, diagrams 4 and 8, especially for final states with $\ell = \mu,
\tau$. The heavy neutrino signal is dominated by diagrams 5 and 6
with $N$ produced on its mass shell, because the $\Gamma_N$ enhancement of the
amplitude partially cancels the mixing angle factor in the decay vertex,
yielding the corresponding branching ratio. It must be remarked that the
$s$-channel $N$ production diagram 7 is negligible (few per mille) when
compared to the  $t$-, $u$-channel diagrams 5 and 6. This behaviour is general,
because the $s$-channel propagator is fixed by the large collider energy and
suppresses the contribution of this diagram, whereas the $t$- and $u$-channel
propagators do not have such suppression.
Since both diagrams 5 and 6 involve an $eNW$
vertex to produce a heavy neutrino, only in the presence of this interaction the
signal is observable. Once the heavy neutrino is produced, it can decay
to $\ell W$ with $\ell = e,\mu,\tau$, being the corresponding branching 
ratios in the ratio $|V_{eN}|^2\, :\, |V_{\mu N}|^2\, :\, |V_{\tau N}|^2$.

In case that the signal is dominated by $t$ and $u$ channel on-shell $N$
production (the only situation in which it is observable), negative electron
polarisation and positive positron polarisation increase its statistical
significance. For the signal contributions alone we have 
$\sigma_{e_R^+ e_L^-} \,:\, \sigma_{e_L^+ e_R^-} = 1200 \,:\, 1$ (for $m_N =
300$ GeV, $V_{eN} = 0.073$), with
$\sigma_{e_R^+ e_R^-} = \sigma_{e_L^+ e_L^-} = 0$. For the SM process,
$\sigma_{e_R^+ e_L^-} \,:\, \sigma_{e_R^+ e_R^-} \,:\,
\sigma_{e_L^+ e_R^-} = 150 \,:\, 7 \,:\, 1$, $\sigma_{e_L^+ e_L^-} = 0$.
In the limit of perfect beam polarisations $P_{e^-} = -1$, $P_{e^+}=1$,
the signal is enhanced with respect to the background by a factor of 1.05 and,
what is more important, the ratio $S/\sqrt B$ increases by a factor of two.
Using right-handed electrons and left-handed positrons decreases the $S/B$
ratio by a factor of 8, and $S/\sqrt B$ by a factor of 50.
On the other hand, if the neutrino does not mix with the electron but mixes with
the muon or tau, the behaviour is the opposite. Since the only contribution
comes from diagram 7 the use of left-handed positrons and
right-handed electrons actually increases the signal, while reducing the SM
cross section for this process \cite{Me}. This case is of limited practical
interest, since for $V_{eN} = 0$ the signal is barely observable. 

We finally point out that the signal cross section exhibits little dependence on
the heavy neutrino mass, except close to the kinematical limit \cite{GZ3}, and
the final results are almost independent of $m_N$ within the range 200-400 GeV
\cite{paper2}. For our calculations we take
$m_N = 300$ GeV. In contrast with what has been claimed in the literature
\cite{ACMV}, we find equal production cross sections for Dirac and Majorana
neutrinos to a very good approximation. The reason is easy to understand:
while in the present case the signal is
strongly dominated by diagrams 5 and 6 (which give equal contributions to the
cross section and do not interfere because light neutrino masses can be safely 
neglected), for a Dirac neutrino only diagram 5 is
present. On the other hand, the width of a Dirac neutrino is one half of the
width of a Majorana neutrino with the same mixing angles \cite{paper2}.

\section{Generation of signals}
\label{sec:4}

The matrix elements for $e^+ e^- \to \ell^- W^+ \nu \to \ell^- q \bar q' \nu$
are calculated using {\tt HELAS} \cite{helas}, including all spin correlations
and finite width effects. We sum SM and heavy neutrino-mediated diagrams at
the amplitude level. The charge conjugate process is included in all our
results unless otherwise noted.
We assume a CM energy of 500 GeV, with electron polarisation $P_{e^-} = -0.8$
and positron polarisation $P_{e^+} = 0.6$.
The luminosity is taken as 345 fb$^{-1}$ per year \cite{lum}.
In our calculations we take into account the effects of
ISR \cite{isr} and beamstrahlung \cite{peskin,BS2}. For the design luminosity
at 500 GeV we use the parameters $\Upsilon = 0.05$, $N = 1.56$ \cite{lum}.
The actual expressions for ISR and beamstrahlung used in our calculation are
collected  in Ref.~\cite{npb}. We also include a beam energy spread of 1\%.

In final states with $\tau$ leptons, we select $\tau$ decays to $\pi$, $\rho$
and $a_1$ mesons (with a combined branching fraction of 55\% \cite{PDB}), in
which a single
$\nu_\tau$ is produced, discarding other hadronic and leptonic decays.
We simulate the $\tau$ decay assuming that the meson and $\tau$ momenta are
collinear (what is a good approximation for high $\tau$ energies) and assigning
a random fraction $x$ of the $\tau$ momentum to the meson, according to the
probability distributions \cite{taudecays}
\begin{equation}
P(x) = 2 (1-x)
\end{equation}
for pions, and
\begin{equation}
P(x) = \frac{2}{2 \zeta^3-4 \zeta^2+1} \left[ (1-2 \zeta^2)-(1-2 \zeta) x
\right]
\end{equation}
for $\rho$ and $a_1$ mesons, where $\zeta = m_{\rho,a_1}^2/m_\tau^2$. We assume
a $\tau$ jet tagging efficiency of 50\%.

We simulate the calorimeter and tracking resolution of the detector by
performing a Gaussian smearing of the energies of electrons, muons 
and jets, using the specifications in Ref.~\cite{tesla2},
\begin{equation}
\frac{\Delta E^e}{E^e} = \frac{10\%}{\sqrt{E^e}} \oplus 1 \% \;, \quad
\frac{\Delta E^\mu}{E^\mu} = 0.02 \% \, E^\mu \;, \quad
\frac{\Delta E^j}{E^j} = \frac{50\%}{\sqrt{E^j}} \oplus 4 \% \;,
\end{equation}
respectively, where the two terms are added in quadrature and the energies 
are in GeV.
We apply kinematical cuts on transverse momenta, $p_T \geq 10$ GeV, and
pseudorapidities $|\eta| \leq 2.5$, the latter corresponding to polar angles
$10^\circ \leq \theta \leq 170^\circ$. To ensure high $\tau$ momenta (so that
the meson resulting from its decay is effectively collinear) we require $p_T
\geq 30$ GeV for $\tau$ jets. We reject events in which the
leptons or jets are not isolated, requiring a ``lego-plot'' separation
$\Delta R = \sqrt{\Delta \eta^2+\Delta \phi^2} \geq 0.4$.
For the Monte Carlo integration in 6-body phase space we use
{\tt RAMBO} \cite{rambo}.

In final states with electrons and muons the light neutrino momentum $p_\nu$ is
determined from the missing transverse and longitudinal momentum of the event
and the requirement that $p_\nu^2 = 0$ 
(despite ISR and beamstrahlung, the missing longitudinal momentum approximates
with a reasonable accuracy the original neutrino momentum).
In final states with $\tau$
leptons, the reconstruction is more involved, due to the additional neutrino
from the $\tau$ decay. We determine the ``first'' neutrino momentum and the
fraction $x$ of the $\tau$ momentum retained by the $\tau$ jet using the
kinematical constraints
\begin{align}
E_W + E_\nu + \frac{1}{x} E_j & = \sqrt s \,, \nonumber \\
\vec p_W + \vec p_\nu + \frac{1}{x}  \vec p_j & = 0 \,, \nonumber \\
p_\nu^2 & = 0 \,,
\end{align}
in obvious notation.
These constraints only hold if ISR and beamstrahlung are ignored, and in the
limit of perfect detector resolution. When solving them for the generated Monte
Carlo events we sometimes obtain $x > 1$ or $x < 0$. In the first case we
arbitrarily set $x=1$, and in the second case we set $x = 0.55$, which is the
average momentum fraction of the $\tau$ jets. With the procedure outlined here,
the reconstructed $\tau$ momentum reproduces with a fair accuracy the original
one, while the obtained $p_\nu$ is often quite different from its original 
value.

\section{Heavy neutrino discovery at ILC}
\label{sec:5}

Following the discussion in Sections \ref{sec:2} and \ref{sec:3} 
we can distinguish two interesting scenarios for our analysis:
({\em i\/}) the heavy neutrino only mixes with the electron; ({\em ii\/}) it
mixes with $e$ and either $\mu$, $\tau$, or both. A third  less interesting
possibility is that
the heavy neutrino does not mix with the electron. We discuss these three cases
in turn.

\subsection{Mixing only with the electron}
\label{sec:5.1}

\begin{figure}[htb]
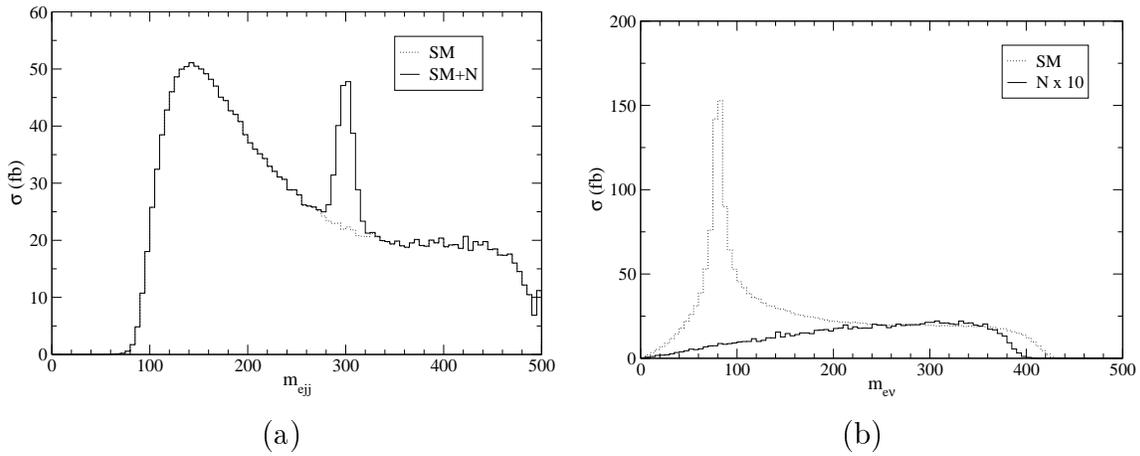

\begin{center}
\begin{tabular}{cc}
\epsfig{file=Figs/mn-e.eps,width=7.3cm,clip=} &
\epsfig{file=Figs/mw-e.eps,width=7.3cm,clip=} \\
(a) & (b)
\end{tabular}
\caption{Kinematical distributions of the $ejj$ invariant mass (a)
and the $e\nu$ invariant mass (b).}
\label{fig:mnmw}
\end{center}
\end{figure}

A heavy neutrino coupling to the electron yields a peak in the
distribution of the $ejj$ invariant mass $m_{ejj}$, plotted in
Fig.~\ref{fig:mnmw}
(a) for $V_{eN} = 0.073$. The solid line corresponds to the SM plus a
300 GeV Majorana neutrino, being the dotted line the SM
prediction. The width of the peak is due
to energy smearing applied in our Monte Carlo and not to the intrinsic neutrino
width $\Gamma_N = 0.14$.

This already striking signal can be enhanced applying a veto cut on the
$e\nu$ invariant mass $m_{e\nu}$, shown in Fig.~\ref{fig:mnmw} (b) for the SM
(dashed line) and the new heavy neutrino signal alone (solid line). The two
contributions have 
been separated for clarity. The SM and SM plus heavy neutrino cross sections are
collected in Table~\ref{tab:cs}, before and after the kinematical cuts
\begin{eqnarray}
& 290 ~\mathrm{GeV} \leq m_{ejj} \leq 310 ~\mathrm{GeV} \,, \nonumber \\
& m_{e\nu} \leq 40 ~\mathrm{GeV} ~\mathrm{or}~ m_{e\nu} \geq 110 ~\mathrm{GeV}
 \,.
\label{ec:cuts}
\end{eqnarray}

\begin{table}[htb]
\begin{center}
\begin{tabular}{lcccc}
& No cuts & $m_{ejj}$ & $m_{e\nu}$ & $m_{ejj}$, $m_{e\nu}$ \\
\hline
SM       & 2253 & 89.1 & 1387 & 53.6 \\
SM + $N$ & 2339 & 173.7 & 1489 & 130.8 
\end{tabular}
\caption{Cross sections (in fb) for $e^+ e^- \to e^\mp W^\pm \nu$ before and
after the kinematical cuts in Eqs.~(\ref{ec:cuts}).}
\label{tab:cs}
\end{center}
\end{table}

The new neutrino is said to be discovered when the excess of events (the
signal $S$) in the peak region amounts to more than 5 standard deviations of
the number of expected events (the background $B$), that is, $S/\sqrt B \geq 5$.
\footnote{It must be noted that the SM cross section at the peak can be
calculated and normalised using the measurements far 
from this region.}
This ratio is larger than $5$ for $V_{eN} \geq 1.2 \times 10^{-2}$,
which is the minimum mixing angle for which a 300 GeV neutrino can be
discovered. If no signal is found, the limit $V_{eN} \leq 6.7 \times
10^{-3}$ can be set at 90\% confidence level (CL), improving the present limit
$V_{eN} \leq 0.073$ by a factor of ten.

\begin{figure}[htb]
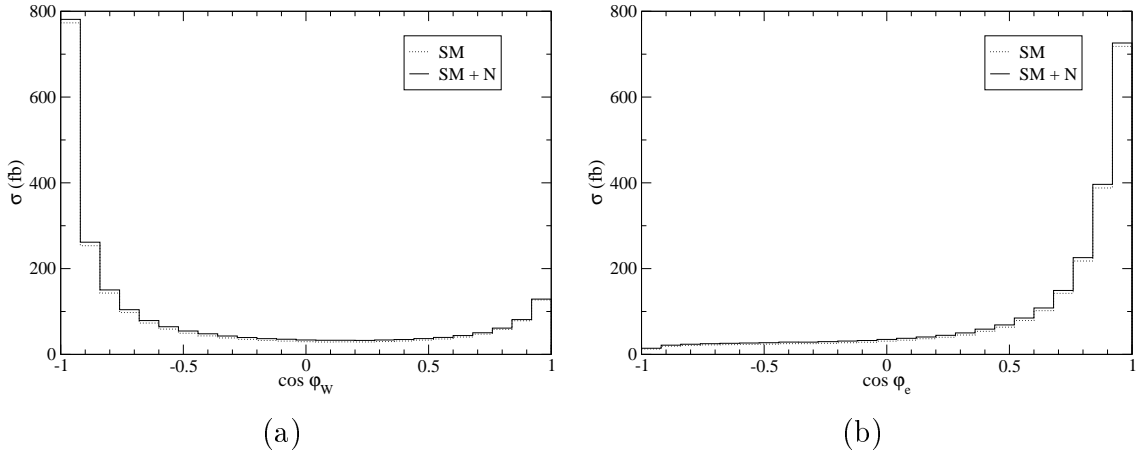

\begin{center}
\begin{tabular}{cc}
\epsfig{file=Figs/cosph-W.eps,width=7.3cm,clip=} &
\epsfig{file=Figs/cosph-e.eps,width=7.3cm,clip=} \\
(a) & (b)
\end{tabular}
\caption{Dependence of the cross section on the angles $\varphi_W$ (a) and
$\varphi_e$ (b), for the SM and the SM plus a heavy Majorana neutrino.}
\label{fig:cosph-We}
\end{center}
\end{figure}

We have also examined the potential of the angular distributions of the
produced $W^\pm$, $e^\mp$ to signal the presence of a heavy neutrino. 
We define the angle $\varphi_W$ as the polar angle between the $W$ and
the electron (in $e^- W^+ \nu$ final states) or positron (for the charge
conjugate process). The angle $\varphi_e$ is defined analogously. 
Their kinematical distributions are shown in Fig.~\ref{fig:cosph-We}. 
From the comparison of these plots with Fig.~\ref{fig:mnmw} (a)
it is apparent that the best kinematical variable to signal the presence of a
heavy neutrino is the $ejj$ invariant mass. For these two angular
distributions the deviation of the SM prediction amounts to
$\chi^2/\mathrm{d.o.f.} \simeq 10000/25$,
$10500/25$, respectively. Dividing the $m_{ejj}$ distribution in 25 bins
for better comparison, the corresponding deviation is
$\chi^2/\mathrm{d.o.f.} \simeq 64000/25$.

The quantitative results obtained here hold for heavy neutrino masses in the
range $200-400$ GeV to a very good approximation \cite{paper2}.
Although for heavier $N$ the
signal cross sections are smaller, the SM background decreases for larger
$m_{ejj}$ as well, as can be seen in Fig.~\ref{fig:mnmw}, and the two effects
compensate. For $m_N > 400$ GeV, the cross sections decrease quickly and thus
the limits obtained for the mixing angles are worse.

\subsection{Mixing with the three charged leptons}
\label{sec:5.3}

In the most general case that a heavy neutrino mixes simultaneously with the
three charged leptons, there may be in principle signals in the $e$, $\mu$,
$\tau$ channels. The three of them must be experimentally analysed in the search
for a heavy neutrino. As we will show in the following, it is possible
that the clearest signals come from the $\mu$ or $\tau$ channels, even despite
the fact that an $eNW$ coupling is necessary to observe them.
To prove it we choose the values $V_{eN} = 0.073/\sqrt 2$,
$V_{\mu N} = 0.098/\sqrt 2$, $V_{\tau N} = 0.13$. These figures are conservative
in the
sense that the heavy neutrino mainly decays to $\tau$ leptons, which are harder
to see experimentally, and the cleanest electron and muon channels are
relatively suppressed. The cross sections after the kinematical cuts in
Eqs.~(\ref{ec:cuts}) can be found in Table~\ref{tab:cs2} for the SM and the SM
plus a heavy neutrino, and for the three modes 
(in the $\tau$ channel we do not
apply the veto cut on $m_{\tau \nu}$). After one year
of running, the heavy neutrino signal could be seen with $11 \sigma$, $24
\sigma$, $24 \sigma$ in the $e$, $\mu$, $\tau$ final states, respectively.

\begin{table}[htb]
\begin{center}
\begin{tabular}{lccc}
& $e$  & $\mu$ & $\tau$ \\
\hline
SM        & 53.0 & 31.8  & 9.7 \\
SM + $N$  & 57.4 & 39.1  & 13.8 
\end{tabular}
\caption{Cross sections (in fb) for $e^+ e^- \to \ell^\mp W^\pm \nu$, for
$\ell=e,\mu,\tau$, including the kinematical cuts in Eqs.~(\ref{ec:cuts}).}
\label{tab:cs2}
\end{center}
\end{table}

For equal couplings $V_{eN} = V_{\mu N}$ the statistical
significance $S/\sqrt B$ of the electron and muon signals is similar. Therefore,
mixing with the muon does not reduce the sensitivity to $V_{eN}$ \cite{paper2}.
This
follows from the fact that for a fixed $V_{eN}$ the $N$ production cross
section is independent of $V_{\mu N}$, while the branching ratios in these two
channels are in the relation $|V_{eN}|^2 \; : \; |V_{\mu N}|^2$. For $V_{\mu N}
= 0$ the charged current decays reduce to $N \to eW$, while for $V_{\mu N}$
larger than $V_{eN}$
 $N \to \mu W$ dominates, being the combined statistical significance of both
channels similar to the one of the electron channel alone for $V_{\mu N} = 0$.
The same argument shows that mixing with the tau lowers the
sensitivity to $V_{eN}$, because the observation of the heavy neutrino in the
$\tau W \nu$ channel is more difficult.

\subsection{Heavy neutrinos not coupling to the electron}
\label{sec:5.4}

We have previously argued that a heavy neutrino signal in the
$\mu$ or $\tau$ channels is observable only if the neutrino also mixes
with the electron. We now quantify this statement. 
We consider a heavy neutrino coupling only to the
muon, with $V_{\mu N} = 0.098$, or only to the tau, with
$V_{\tau N} = 0.13$. The beam polarisations $P_{e^-}=0.8$,
$P_{e^+}=-0.6$, opposite to the previous ones, are used to enhance the signal
and reduce the SM background. The cross sections for the SM and SM plus a heavy
neutrino are shown in Table~\ref{tab:cs3} for these two cases. For $N$ mixing
only with the muon, the statistical significance of the signal is
$S/\sqrt B = 1.85$  for one year of running, and 7.3 years would be necessary
to observe a $5 \sigma$ deviation. If the heavy neutrino only
mixes with the $\tau$, the statistical significance is only
$S/\sqrt B = 0.76$, in which case a luminosity 43 times larger is required to
achieve a $5\sigma$ evidence.

\begin{table}[htb]
\begin{center}
\begin{tabular}{lcc}
& $N-\mu$ & $N-\tau$ \\
\hline
SM        & 1.10 & 0.389 \\
SM + $N$  & 1.20 & 0.414
\end{tabular}
\caption{$e^+ e^- \to \ell^\mp W^\pm \nu$ cross sections (in fb) for 
a heavy neutrino coupling only to the muon (first column, $\ell=\mu$)
or coupling only to the tau (second column, $\ell=\tau$), 
and the kinematical cuts in Eqs.~(\ref{ec:cuts}).}
\label{tab:cs3}
\end{center}
\end{table}

\section{Conclusions}
\label{sec:6}

Heavy neutrinos with masses near the electroweak scale and large mixing angles
$\sim 0.1 - 0.01$ with the SM leptons are observable at ILC if they exist. Here
we have studied the ILC potential for their detection  in the process
$e^+ e^- \to \ell W \nu$, taking into account the SM background and the effects
of ISR and beamstrahlung, paying special attention to the relevance of the final
state lepton flavour and initial beam polarisation. Using a parton simulation it
has been shown that it is possible to observe a heavy neutrino signal in this
final state if it has a mixing with the electron $V_{eN} \gtrsim 10^{-2}$. 
Although a mixing with the electron of this size is necessary to observe a heavy
neutrino at ILC, the signal may be more visible in the muon or tau channel if it
also has a relatively large coupling to them. The production cross sections and
then the discovery limits do not depend on the Dirac or Majorana nature of the
heavy neutrino. 

These non-decoupled heavy neutrinos do not explain the observed light neutrino
masses nor the baryon asymmetry of the universe. In this sense this search for
heavy neutrinos at large lepton colliders is complementary to the joint
experimental effort for determining the light neutrino properties, and in
particular the neutrino mixing matrix \cite{APS}. We could also know about
non-decoupled heavy neutrinos if the MNS matrix was found to be non-unitary 
or CP violation beyond the allowed limits for the minimal SM extension with
light Dirac or Majorana masses was measured \cite{BGHSZ}. (CP violation is
unobservable in $e^+ e^- \to N \nu \to \ell W \nu$ at ILC because the possible
effects and statistics are small.) Other signals of heavy neutrinos like pair
production \cite{CKK} are suppressed by extra powers of the small mixing angles
and the center of mass energy threshold. In models with extra matter or
interactions further new physics signatures are possible \cite{LLM}. For
example, in left-right symmetric models the new gauge bosons can mediate heavy
neutrino single and pair production \cite{clicreview}, but we assume that they
are too heavy to produce a signal at ILC. At any rate, we are interested in the
production of heavy neutrinos having relatively large mixings with
the SM fermions.

Finally, other future experiments might exhibit indirect signals of these
non-decoupled heavy neutrinos, or limit the possibility of their observation at
large colliders. If no deviation from the SM predictions for the invisible $Z$
width is observed in the Giga$Z$ option of an $e^+ e^-$ collider \cite{tdr}, 
the bound on their mixing with the  light leptons will be reduced by
more than one order of magnitude. Analogously, future improvements on the limits
on flavour violating processes like $\mu \to e \gamma$ \cite{MEG} or $\mu - e$
conversion \cite{MECO} by several orders of magnitude would also reduce the
corresponding bounds on neutrino mixing by more than one order of magnitude,
implying a reduction of the possible signals at ILC or requiring a more
delicate fine-tuning. In all cases the eventual improvement on mixing angle
constraints is comparable to the ILC potential.

\vspace{1cm}
\noindent
{\Large \bf Acknowledgements}

\vspace{0.4cm} \noindent
F.A. thanks J. Bernab\'eu, A. Bueno, J. Wudka and M. Zra{\/l}ek for discussions.
This work has been supported in part by MEC and FEDER Grant No.
FPA2003-09298-C02-01, by Junta de Andaluc{\'\i}a Group FQM 101,
by FCT through projects POCTI/FNU/44409/ 2002,
CFTP-FCT UNIT 777 and grant SFRH/BPD/12603/2003, and by the European
Community's Human Potential Programme under contract HPRN-CT-2000-00149 Physics
at Colliders.

\end{document}